\documentclass[a4paper,12pt]{amsart}
\usepackage{epsfig}
\usepackage{amsfonts,amsmath,latexsym,amssymb,txfonts,bm}
\usepackage[american]{babel}
\newcommand\beq{\begin{eqnarray}}
\newcommand\eeq{\end{eqnarray}}
\newcommand\be{\begin{equation}}
\newcommand\ee{\end{equation}}
\newcommand\nn{\nonumber}

\newcommand{\cal}{\mathcal}
\newcommand{\om}{{\omega}}\newcommand{\Ref}[1]{(\ref{#1})}
\begin{document}

\title{Casimir energies in spherically symmetric background potentials revisited}

\author{Matthew Beauregard}
\address{Department of Mathematics, Stephen F. Austin State University, Nacogdoches, TX, 75962}
\email{beauregama@sfasu.edu}
\author{Michael Bordag}\thanks{MB and KK acknowledge support by the DFG under contract number BO 1112/22-1}
\address{Universit{\"a}t Leipzig, Institute for Theoretical Physics, Postfach 100920, 04009 Leipzig, Germany}
\email{Michael.Bordag@uni-leipzig.de}
\author{Klaus Kirsten}
\address{Department of Mathematics, Baylor University, Waco, TX 76798, USA}
\email{Klaus\textunderscore Kirsten@baylor.edu}

\begin{abstract}
In this paper we reconsider the formulation for the computation of the Casimir energy in spherically symmetric background potentials.
Compared to the previous analysis, the technicalities are much easier to handle and final answers are surprisingly simple.
\end{abstract}
\date{\today}
\maketitle

\section{Introduction}
Quantum corrections to classical solutions that are not constant are important in a variety of areas in modern physics.
Inhomogeneous classical configurations are, for example, the rule when considering monopoles \cite{poly74-20-194,hoof74-79-276},
sphalerons \cite{klin84-30-2212} and electroweak Skyrmions \cite{adki83-228-552,ambj85-256-434,eila86-56-1331,frie77-15-1694,gips84-231-365,gips81-183-524,frie77-16-1096,skyr61-260-127,skyr62-31-556}.
Given this importance, many efforts have been made to develop a combination of analytical and numerical techniques for the computation
of those quantum corrections \cite{baac90-47-263,baac93-48-5648,bord00-61-085008,bord96-53-5753,brah94-49-4094,dunn05-94-072001,dunn06-74-085025,dunn08-77-045004,dunn11-83-105013,dunn09-42-075402,grah04-677-379,grah02-645-49,lee94-49-4101,rahi09-80-085021}.

For spherically symmetric classical backgrounds a standard approach is to use quantum mechanical scattering data for the analysis. Whereas in $1+1$ dimensions the associated analysis gives results that can be very easily handled numerically \cite{boch92-46-5550,bord95-28-755},
in higher dimensions additional summations over the angular moments complicate the situation \cite{bord96-53-5753,grah02-645-49}.
An essential element of such calculations is the method of momentum integration, for which there are two possibilities.
{\it First}, one integrates over real frequencies, or, {\it second}, one turns the integration path, $\om=i\xi$, and integrates over imaginary frequencies $\xi$. The second way is commonly used. Here, the integration is over the logarithm of the Jost functions (see eq.\Ref{10} below). This is a smooth function and after the necessary subtractions, one comes to a convergent integration which can be well handled numerically. Also, as a technical advantage, bound states and zero modes are included automatically. In the first way, represented in eq. \Ref{11} below, where the integration is over real frequencies,
the integrand has generically an oscillating behaviour resulting from the scattering phase shifts. This has the potential to make numerical integrations more difficult.
Also, in this method one has to bother about the bound states and to include them explicitly (see eq.\Ref{11}). Furthermore, when integrating by parts, one has to pay attention to boundary contributions
following with Levinson's theorem.

%
%
%
%
%
%

Another somewhat frustrating aspect of this calculation is that from the outset, divergent contributions are known through heat kernel coefficients
but no use of this knowledge is being made. It is this aspect of the analysis that has been improved in our current work. It is shown
that sums over phase shifts for large moments can be expressed as heat kernel coefficients. This makes the subtraction and renormalization
of divergencies very straightforward and the remaining incumbent numerical approximations are much easier than previously.

The article is organized as follows. First, in Section 2, we present some basic formulas for the zeta function analysis of vacuum energies
as obtained earlier \cite{bord96-53-5753}. In Section 3, we then establish the indicated relation between a sum over phase shifts and corresponding
heat kernel coefficients that will lead to an extremely simple expression for the renormalized Casimir energy. This expression is then used
in Sections 4 and 5 to present examples, namely step potentials and delta potentials, in two and three dimensions. The Conclusions
summarize the most important aspects of our work and point to possible future activities along the line presented here.

\section{Zeta functions in terms of Jost function and phase shift}
In this section we provide the essential formulas for the zeta function computation in the relevant context. The exposition follows \cite{bord96-53-5753}.

The Casimir energy of a massive non-interacting scalar field in a spherically symmetric background field is obtained from the
eigenvalue equation
\beq
(-\Delta + V(r)) \phi_j (\vec x) = \lambda_j^2 \phi_j (\vec x) . \label{1}
\eeq
In the zeta function scheme the Casimir energy for a field of mass $m$ follows from the zeta function
\beq
\zeta (s) = \sum_j ( \lambda_j^2 + m^2 )^{-s} \label{2}
\eeq
as
\beq
\left. E_{Cas}(s) = \frac 1 2 \zeta \left( s-\frac 1 2 \right) \mu^{2s} \right|_{s=0}, \label{3}
\eeq
where $\mu$ is an arbitrary mass parameter. Imposing the renormalization condition that the quantum
fluctuation of a quantum field should die out as the mass of the field tends to infinity, namely
\beq
\lim_{m\to\infty} E_{Cas}^{(ren)}(s) =0, \label{4}
\eeq
leads to the subtraction of
\beq
E_{Cas}^{div}(s) &=& -\frac{m^4}{8 \sqrt {\pi}} \left( \frac 1 s + \ln \left[ \frac{4\mu^2}{m^2}\right]
 - \frac 1 2 \right) a_0 - \frac{m^3} 3 a_{1/2}\nn\\
 & &+ \frac{m^2}{4 \sqrt \pi} \left( \frac 1 s + \ln \left[ \frac{4\mu^2} {m^2} \right] -1\right) a_1 + \frac 1 2 m a_{3/2} \label{5}\\
 & &- \frac 1 {4 \sqrt \pi} \left( \frac 1 s + \ln \left[ \frac{4\mu^2}{m^2} \right] - 2 \right) a_2.\nn
\eeq
Here, $a_i$, $i=0,1/2,1,3/2,2$, are the heat-kernel coefficients associated with the operator $-\Delta + V(r)$, and describe the small-$t$ asymptotics of the heat kernel
\beq
K(t) = \sum_j e^{-\lambda_j^2 t} \raisebox{-4pt}{$\sim \atop {t\to 0}$} \ \ \frac 1 {(4\pi t)^{3/2} } \sum_{i=0,1/2,1,...}^\infty a_i \, t^i .\label{6}
\eeq
Although half-integer heat kernel coefficients typically occur only on manifolds {\it with} a boundary, however, they are included here since
potentials which are not differentiable or are singular at some points make the inclusion necessary.

We therefore have
\beq
E_{Cas}^{(ren)}(s) = E_{Cas}(s) - E_{Cas}^{div}(s) . \label{7}
\eeq
The remaining task is to find a representation of the zeta function valid in the neighborhood of $s=-1/2$, respectively of the vacuum energy $E_{Cas}^{(ren)}(s)$ in $s=0$.
In fact, the main difficulties of the first approach are hidden here. We know, that the limit $s\to0$ in \Ref{7} gives a finite result. However, both quantities there are divergent. A method to handle this problem was developed in \cite{bord96-53-5753}. There the vacuum energy was represented in terms of the Jost function. To derive these
representations note that, after separation of variables, the relevant radial eigenvalue equation is
\beq
\left[ \frac{d^2}{dr^2} - \frac{\ell (\ell +1) } {r^2} - V(r) + p^2 \right] \psi_{\ell , p} (r) = 0 , \label{8}
\eeq
with eigenvalue $p^2$ and angular momentum $\ell$, $\ell \in {\mathbb N}_0$. Except for very simple potentials, the eigenvalues $p^2$
and eigenfunctions $\psi_{\ell , p}(r)$ will not be known explicitly, but the needed information can be found from scattering theory.

The asymptotic $r\to\infty$ behavior of the eigenfunctions $\psi_{\ell ,p}(r)$ is described by the so-called Jost function \cite{tayl72b}. It is
defined based upon the regular solution $\phi_{\ell ,p}(r)$, which is a normalized eigenfunction such that as $r\to 0$
\beq
\phi_{\ell ,p} (r) \sim \hat j _\ell (pr) \nn
\eeq
with the spherical Bessel function $\hat j _\ell (z)$. This function has the asymptotic behavior, for $r\to\infty$,
%
\beq
\phi _{\ell , p}(r) \sim \frac i 2 \left[ f_\ell (p) \hat h _\ell^- (pr) - f_\ell ^* (p) \hat h _{\ell } ^+ (pr) \right] , \label{9}
\eeq
with the Riccati-Hankel function $\hat h_\ell^\pm (z)$ and the coefficients $f_\ell (p)$ are the Jost function and its complex conjugate. This equation says nothing but that outside of
the support of the potential (which for the moment can be assumed to be compact), the solution can be written as a linear combination of two linearly independent solutions of the free differential
equation, namely (\ref{8}) with $V(r)=0$.

Putting the system temporarily in a big sphere of radius $R$, imposing Dirichlet boundary conditions at $r=R$,
eigenvalues of (\ref{8}) are determined by
\beq
f_\ell (p) \hat h_\ell ^- (pr) - f_\ell ^* (p) \hat h _\ell ^+ (pr) =0 . \label{9a}
\eeq
This transcendental equation is then used together with the residue theorem to write down a complex contour
integral representation of the zeta function, a standard technique by now \cite{kirs02b}. Subtracting off the free Minkowski contribution one obtains \cite{bord96-53-5753}
\beq
\zeta (s) = \frac{\sin \pi s} \pi \sum_{\ell =0}^\infty (2\ell +1) \int\limits_m^\infty dk (k^2-m^2)^{-s} \frac d {dk}
\ln f_\ell (ik) .\label{10}
\eeq
The analytical continuation of the zeta function or the vacuum energy needed in eq. \Ref{7}  is then obtained by subtracting
and adding the leading terms in the uniform asymptotic expansion of the Jost function obtained from the Lippmann-Schwinger
equation \cite{tayl72b}. Technicalities are somewhat involved, especially for spinor fields \cite{bord99-60-105019}, and the representation obtained for the Casimir energy has a somewhat
unpleasant appearance, though a numerical approximation is possible \cite{bord96-53-5753}.

Notice, that in this approach the momentum integration (over imaginary frequencies) is done first and the orbital momentum sum is performed afterwards.

Instead of using the Jost function as the basic quantity for the representation, one might equally well use the scattering phase
$\delta_\ell (q)$. Denoting by $-\kappa_{n,\ell}^2$ the energies of the bound states, the dispersion relation for the Jost function \cite{tayl72b}
\beq
f_\ell (ik) = \prod _n \left( 1-\frac{\kappa_{n,\ell}^2}{k^2}\right) \exp \left( - \frac  2 \pi \int\limits_0^\infty
\frac{dq \,\, q}{q^2+k^2} \delta_\ell (q)\right),\label{disp}\eeq
leads to the pertinent representation
\beq
\zeta (s) &=& \sum_{\ell =0}^\infty \left( 2 \ell + 1 \right) \left\{
- \sum_n \left( m^{-2s} - \left(m^2-\kappa_{n,\ell}^2\right)^{-s} \right)
+ \frac {2s} \pi \int\limits_0^\infty dq \,\frac{q}{ (q^2+m^2)^{s+1}} \delta_\ell (q) \right\}.\label{11}
\eeq
Also here, one now subtracts and adds a suitable asymptotic expansion of the phase shifts to obtain a valid representation for the Casimir energy.

The novel point in our current approach is to interchange summation and integration and to analyze the sum over phase shifts further. This sum is
known to converge as long as $V(r) \sim 1/r^{3+\epsilon}$, $\epsilon >0$, for $r\to\infty$ (see, for example, \S 123 in \cite{land77b}). As we will see, choosing this prescription to proceed no
computation of asymptotic expansions will be necessary and the Casimir energy representation becomes as compact as it could possibly be!

\section{Summing phase shifts}
In order to simplify the presentation let us assume that there are no bound states. Then
\beq
\zeta (s) = \frac {2s} \pi \sum_{\ell =0}^\infty (2 \ell +1) \int\limits_0^\infty dq \,\, \frac{q }{(q^2+m^2)^{s+1}} \, \delta_\ell (q) .\label{12}
\eeq
Interchanging summation and integration, with the notation
\beq
\delta (q) = \sum_{\ell =0}^\infty (2\ell +1) \delta_\ell (q) , \label{13}
\eeq
one finds
\beq
\zeta (s) = \frac{2s} \pi \int\limits_0^\infty dq \,\, q (q^2+m^2)^{-s-1} \delta (q ). \label{14}
\eeq
As indicated, the quantity $\delta (q)$ is well defined for suitable potentials and known properties
of the zeta function suggest, that eq. (\ref{14}) is well defined for $\Re s >3/2$ because $s=3/2$ is the location of the right-most pole.

Known properties of the phase shift $\delta_\ell (q)$ show that $\delta (q)$ remains finite as $q\to 0$. The singularities of the zeta function,
located in general at $s=3/2,1,1/2,-(2k+1)/2$, $k\in{\mathbb N}_0$, must therefore be a consequence of the large-$q$ expansion of $\delta (q)$.
That expansion is independent of $m$ and using inverse Mellin-transforms \cite{titc48b}, with $\zeta_0 (s)$ denoting the zeta function for the massless case, we
find
\beq \delta (q) = - \frac \pi {2\pi i} \int\limits_{c-i\infty}^{c+i\infty} ds \,\, q^{-s} \frac{\zeta_0 \left( - \frac s 2 \right)} s \label{15}
\eeq
for $\Re c > -3/2$. Shifting the contour to the right, we pick up the large-$q$ expansion of $\delta (q)$. The precise form depends on the
location of the poles of $\zeta_0 (-s/2)$ and of its residues, furthermore its value at $s=0$ enters.
All this information is encoded in the heat kernel
coefficients which are known for a wide class of potentials $V(r)$.

In detail, the residue theorem shows that the large-$q$ behavior reads
\beq
\delta (q) \sim \pi \left( \frac{4q^3}{3 \sqrt \pi} a_0 + q^2 a_{1/2} + \frac{2q} {\sqrt \pi} a_1 + a_{3/2}
+ \frac 1 {q\sqrt \pi} a_2 + ... \right) . \label{16}
\eeq
To construct the analytical continuation of eq. (\ref{14}) to $s=-1/2$, one subtracts the above right hand side from $\delta (q)$ and
adds it back. For the Casimir energy this naturally leads to a splitting in two terms, namely
\beq
E_{Cas} = E_f + E_{as} , \label{17}
\eeq
where
\beq
E_f &=& - \frac 1 {2\pi} \int\limits_0^\infty dq \,\, \frac q {\sqrt{q^2+m^2}} \left\{ \delta (q) - \frac{4\sqrt\pi} 3 a_0 q^3 - \pi a_{1/2} q^2
- 2 \sqrt \pi a_1 q - \pi a_{3/2} - \sqrt \pi a_2 \frac 1 q\right\} \nn
\eeq
and
\beq
E_{as} &=& - \frac{1-2s} {2\pi} \mu^{2s} \left\{ \frac{4\sqrt \pi} 3 a_0 \int\limits_0^\infty dq \,\, \frac{q^4}{(q^2+m^2)^{1/2+s}} + \pi a_{1/2}
\int\limits_0^\infty dq \,\, \frac{q^3}{(q^2+m^2)^{1/2+s} }\right.\nn\\
& &\left. +2 \sqrt \pi a_1 \int\limits_0^\infty dq \,\, \frac{q^2}{(q^2+m^2)^{1/2+s}} + \pi a_{3/2}
\int\limits_0^\infty dq \,\, \frac{q}{(q^2+m^2)^{1/2+s}} + \sqrt \pi a_2
\int\limits_0^\infty dq \,\, \frac{1}{(q^2+m^2)^{1/2+s}}    \right\}\nn\\
&=& - (1-2s) \mu^{2s} \left\{ \frac{m^{4-2s} \Gamma (s-2)}{4 \Gamma \left( s + \frac 1 2 \right)} a_0 + \frac{m^{3-2s}}{3-8s+4s^2} a_{1/2}
+\frac{m^{2-2s} \Gamma (s-1)} {4 \Gamma \left( s+\frac 1 2 \right)} a_1 \right.\nn\\
& &\left. \hspace{2.0cm} + \frac{m^{1-2s}}{2 (2s-1)} a_{3/2} + \frac{m^{-2s} \Gamma (s) } {4 \Gamma \left( s + \frac 1 2 \right)} a_2 \right\} \nn\\
&=& E_{Cas}^{div}+{\cal O} (s) ,\nn\eeq
where in the last step an expansion about $s=0$ has been performed. We thus found
\be
E_{Cas}^{(ren)}=- \frac 1 {2\pi} \int\limits_0^\infty dq \,\, \frac q {\sqrt{q^2+m^2}} \ \delta_{\rm subtr}(q)
\label{18_0}\ee
with the subtracted phase shift
\be \delta_{\rm subtr}(q)= \delta (q) - \frac{4\sqrt\pi} 3 a_0 q^3 - \pi a_{1/2} q^2- 2 \sqrt \pi a_1 q - \pi a_{3/2} - \sqrt \pi a_2 \frac 1 q .\label{18}
\ee
The merit of this formula as compared with the other approach is that here the analytic continuation in the regularization parameter is done. Now, the remaining frequency integration is convergent and directly suited for numerical evaluation.

This simple formula is the starting point for the analysis of examples in three dimensions. Before presenting these, let us briefly describe the modifications necessary when two spatial dimensions are considered.
In two dimensions, the radial equation one obtains after separation of variables is
\beq
\left( \frac{d^2}{dr^2} - \frac{m^2-\frac 1 4} {r^2} - V(r)+p^2 \right) \psi_{m,p} (r) =0.\label{19}
\eeq
The corresponding Jost function follows from
\beq
\phi _{m,p} (r) \sim \frac i 2 \left[ f_m (p) \hat h^- _{m-1/2} (pr) - f_m^* (p) \hat h^+_{m-1/2} (pr) \right],\label{20}
\eeq
where $\phi_{m,p} (r)$ is the regular solution behaving, as $r\to 0$, like
\beq
\phi_{m,p} (r) \sim \hat j _{m-1/2} (pr).\nn
\eeq
From the dispersion relation (\ref{disp}), one finds (\ref{11}) with a change in the degeneracy, which is two
for $m>0$ and one for $m=0$. Now (\ref{14}) is immediate, with
\beq
\delta (q) = \delta_0 (q) + 2 \sum_{m=1}^\infty \delta_m (q) . \label{21}
\eeq
Again from (\ref{15}), the large-$q$ asymptotics follows
\beq
\delta (q) \sim \pi a_0 q^2 + 2 \sqrt \pi a_{1/2} q + \pi a_1 + \sqrt \pi a_{3/2} \frac 1 q + ... \label{22}
\eeq
Splitting the Casimir energy as in (\ref{17}), once again $E_{as} = E_{Cas}^{div}$, and
\beq
E_{Cas}^{(ren)} = - \frac 1 {2\pi} \int\limits_0^\infty dq \frac q {\sqrt{q^2+m^2}} \left\{
\delta (q) - \pi a_0 q^2 - 2 \sqrt \pi a_{1/2} q - \pi a_1 - \sqrt \pi a_{3/2} \frac 1 q\right\}. \label{23}
\eeq
We are now in the position to easily study examples.
\section{Casimir energy for several examples in three dimensions}
A possible way to find the phase shift $\delta_{\ell}(q)$ is to use its relation with the Jost function,
namely \cite{tayl72b}
\beq
\frac{f_\ell (p)}{f_\ell^* (p)} = e^{-2i\delta_\ell (p)} , \label{24}
\eeq
such that
\beq
\delta_\ell (p) = - \arctan \frac{\Im f_\ell (p)}{\Re f_\ell (p) } .\label{25}
\eeq
So once the Jost function is known, the phase shift can be found.

Consider a potential with compact support, say $V(r)=0$ for $r\geq R$ and let $u_{\ell ,p} (r)$ denote the regular solution
inside the support of the potential. The complete regular solution then reads
\beq
\phi_{\ell ,p} (r) = u_{\ell ,p} (r) \Theta (R-r) + \frac i 2 \left[ f_{\ell } (p) \hat h _\ell^- (pr) -
f_\ell ^* (p) \hat h ^+_\ell (pr) \right] \Theta (r-R).\label{26}
\eeq
Imposing continuity of $\phi_{\ell ,p}(r)$ and its derivative at $r=R$, one gets
\beq
f_\ell (p) = - \frac 1 p \left( p u_{\ell ,p}(R) {(\hat h ^+_\ell)}' (pR) - u'_{\ell,p} (R) \hat h_ \ell^+ (pR) \right),\label{27}
\eeq
where the Wronskian of $\hat h^\pm_\ell$ was also used \cite{grad65b}. So once $u_{\ell ,p}(r)$ is known for a given potential (with at worst finite
jumps), the Jost function
and thus phase shift is known analytically as well.

\subsection{Step potential}
\begin{figure}[h]
\begin{center}
\includegraphics[scale=.5]{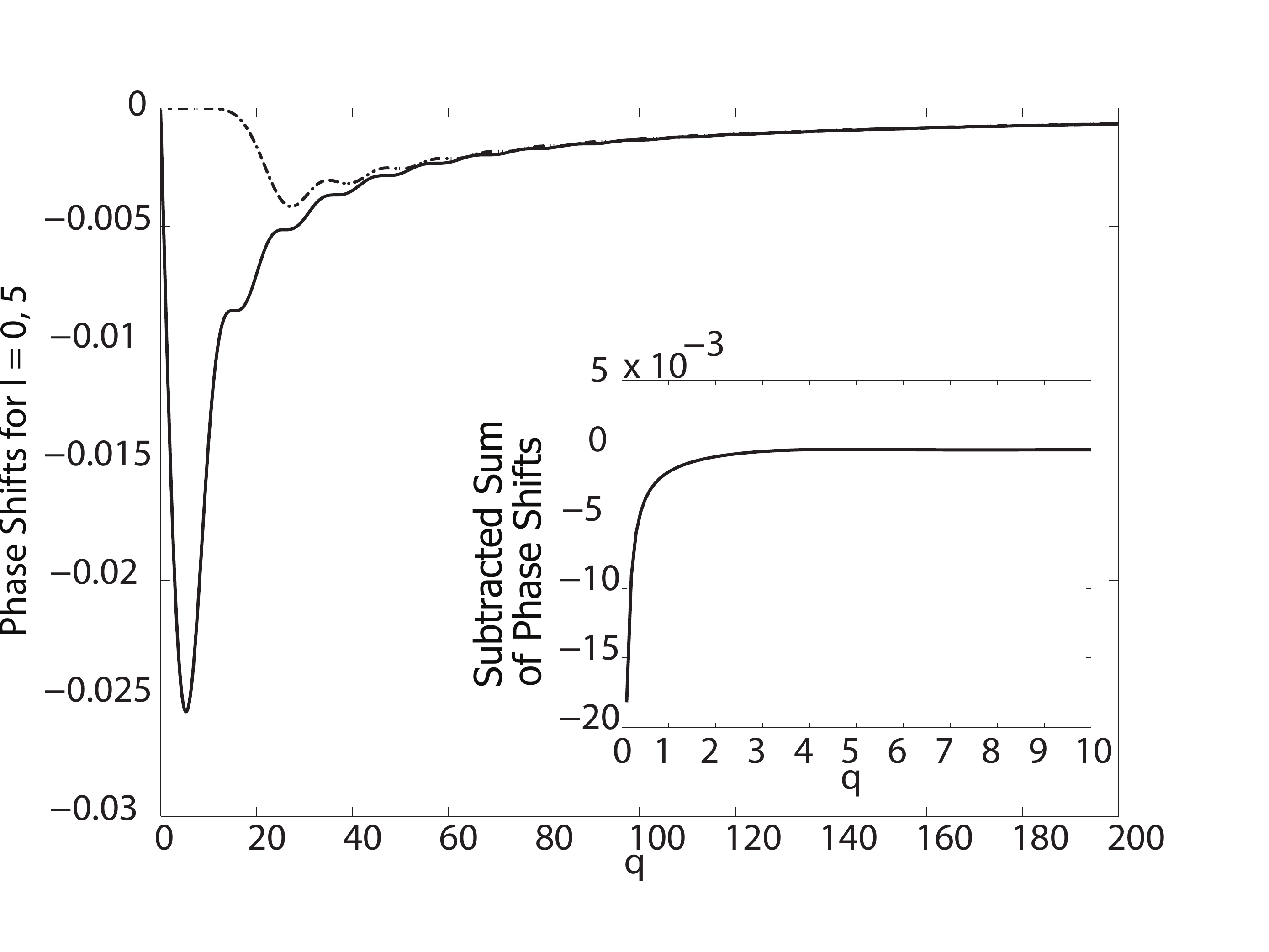}
\caption{Phase shifts for orbital momenta of $\ell=0$ (solid) and $\ell=5$ (dashed-dot) and the subtracted phase shift (inset) for the 3-dimensional case of heaviside potentials for $m=1, V_0=.9, R=1$ as a function of the momentum $q$.}
\label{fig11}
\end{center}
\end{figure}

Consider the example where $V_0 >0$, and
\beq
V(r) = \left\{ \begin{array}{ll}
V_0 & \mbox{for }r\leq R\\
0 & \mbox{for }r>R.\end{array}\right.\label{28}
\eeq
The solution to the eigenvalue problem (\ref{8}) follows from the free solution by replacing $p$ with $q=\sqrt{p^2-V_0}$. Normalizing
correctly, this shows
\beq
u_{\ell ,p} (r) = \left( \frac p q \right)^{\ell +1} \hat j _\ell (qr) \label{29}
\eeq
and thus
\beq
f_\ell (p) = - \left( \frac p q \right)^{\ell +1} \left[ \hat j _\ell (qR) {(\hat h^+_\ell )}' (pR) - \frac q p
\hat j ' _\ell (qR) \hat h ^+ _\ell (pR) \right] . \label{30}
\eeq
Expressing this in terms of Bessel functions of the first and second kind using the relations \cite{abra70b}
\beq
\hat j _\ell (z) = \sqrt{\frac{\pi z} 2 } J_{\ell +1/2} (z) , \quad
\hat h^+ _\ell (z) = i \sqrt{\frac{\pi z} 2} H_{\ell +1/2} ^{(1)} (z), \quad H_{\ell +1/2} ^{(1)} (z) = J_{\ell +1/2} (z) + i N_{\ell +1/2} (z),\nn
\eeq
this reads
\beq
f_\ell (p) &=& \frac{\pi R} 2 \left( \frac p q \right)^{\ell +1/2} \left( i \left[
q J_{\ell +1/2} (pR) J_{\ell +1/2} ' (qR) - p J_{\ell +1/2} (qR) J_{\ell +1/2} ' (pR) \right] \right.\nn\\
& &\left.+ p J_{\ell +1/2} (qR) N_{\ell +1/2} ' (pR) - q N_{\ell +1/2} (pR) J_{\ell +1/2} ' (qR) \right). \label{31}
\eeq
We can now easily read off real and imaginary part of the Jost function to express the phase shift as
\beq
\delta_\ell (p) = - \arctan \frac{ q J_{\ell +1/2} (pR) J_{\ell +1/2} ' (qR) - p J_{\ell +1/2} (qR) J_{\ell +1/2} ' (pR) }
                      {p J_{\ell +1/2} (qR) N_{\ell +1/2} ' (pR) - q N_{\ell +1/2} (pR) J_{\ell +1/2} ' (qR) } .\label{32}
\eeq
To find the Casimir energy for this example from eq. (\ref{18}) we also need the corresponding heat kernel coefficients. Given
the potential has a jump one should refer to the results for non-smooth potentials as given in \cite{gilk01-601-125}. One finds
\beq
a_0 = a_{1/2} = a_{3/2} =0, \quad a_1 = - \frac{R^3 V_0} {6 \sqrt \pi} , \quad a_2 = \frac{R^3 V_0^2}{12 \sqrt \pi} .\label{33}
\eeq
These formulas allow to define the subtracted phase shift \Ref{18}. It is shown as a function of q in Fig. \ref{fig11}.
It is seen that the phase shifts for the individual orbital momenta have some oscillations, however weak in this case, and that the subtracted phase shift is smooth.

\begin{figure}[h]
\begin{center}
\includegraphics{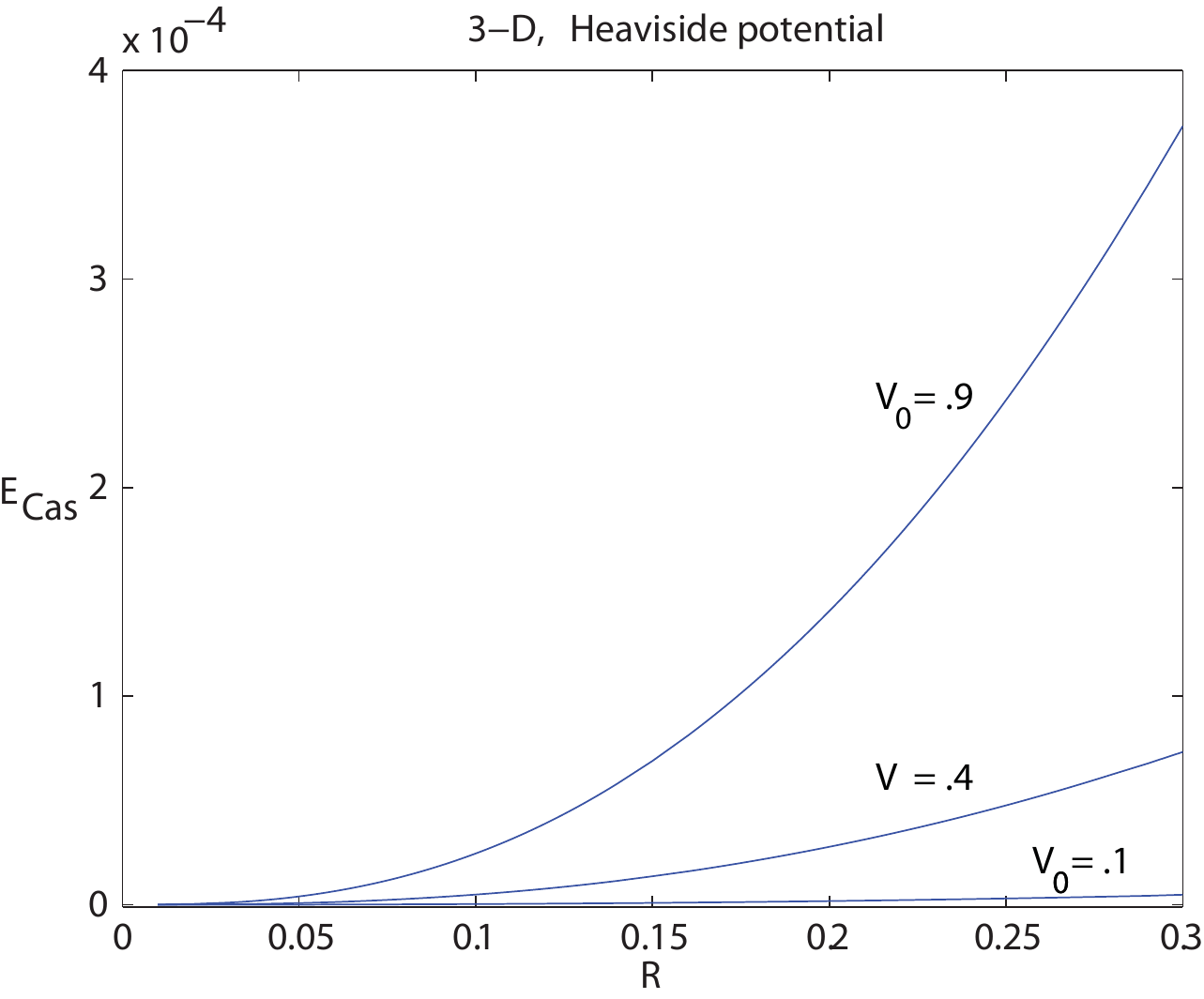}
\caption{The Casimir energy for the 3-dimensional case for heaviside potentials for $m=1$}
\label{fig1}
\end{center}
\end{figure}

Finally, carrying out the integration in \Ref{18_0} numerically one comes to $E_{Cas}^{(ren)}$. For several values of $V_0$ the results are given in Fig. \ref{fig1}  for $m=1$.

\subsection{Delta function potential}
\begin{figure}[h]
\begin{center}
\includegraphics[scale=.5]{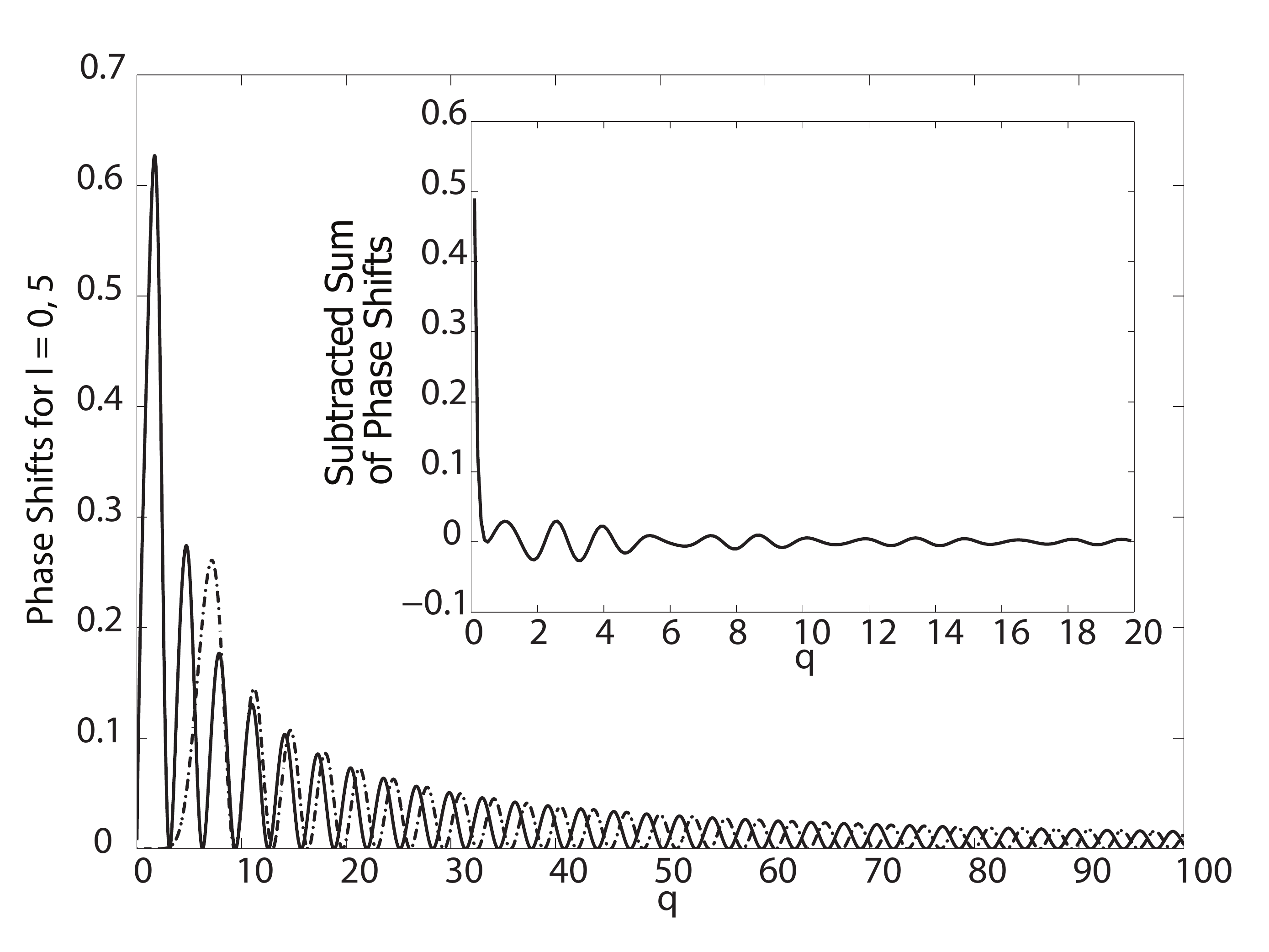}
\caption{Phase shifts an orbital momenta of $\ell=0$ (solid) and $\ell=5$ (dashed-dot) and the subtracted phase shift (inset) for the 3-dimensional case of delta function potentials for $R=1$ as function of the momentum $q$.}
\label{fig22}
\end{center}
\end{figure}

Let us next consider the example
\beq
V(r) = \frac \alpha R \delta (r-R) . \label{34}
\eeq
The solution for $r<R$ then simply is the free solution,
\beq
u_{\ell ,p} (r) = \hat j _\ell (pr) .\nn
\eeq
For this type of singular potential the appropriate boundary conditions are continuity of $\phi_{\ell ,p} (r)$ at $r=R$, so
\beq
\lim_{r\to R^+} \phi_{\ell , p} (r) = \lim_{r\to R^-} \phi_{\ell ,p} (r) , \nn
\eeq
and a jump in the derivative,
\beq
\lim_{r\to R^+} \phi_{\ell ,p} ' (r) = \lim_{r\to R^-} \phi_{\ell ,p} ' (r) + \frac \alpha R \phi_{\ell ,p} (R) .\label{35}
\eeq
From (\ref{26}) one then finds
\beq
f_{\ell } (p) &=& 1 + \frac {i\pi\alpha } 2  H_{\ell +1/2} ^{(1)} (pR) J_{\ell +1/2} (pR) \nn\\
&=& 1 + \frac{i\pi \alpha} 2 J_{\ell +1/2} ^2 (pR) - \frac{\pi \alpha } 2 J_{\ell +1/2} (pR) N_{\ell +1/2} (pR) . \label{36}
\eeq
So
\beq
\delta_\ell (p) = - \arctan \frac{ \frac {\pi \alpha } 2  J_{\ell +1/2} ^2 (pR) } {1-\frac{\pi \alpha } 2 J_{\ell +1/2} (pR)
N_{\ell +1/2} (pR) } . \label{37}
\eeq
The needed leading heat kernel coefficients can once again be found from \cite{gilk01-601-125} and they are
\beq
a_0 =a_{1/2} =0, \quad a_1 = -\frac{\alpha R} {2\sqrt \pi} , \quad a_{3/2} = \frac{\alpha ^2} 8 , \quad a_2 = - \frac{\alpha ^3}
{12 \sqrt \pi R} .\label{38}
\eeq
In this case the phase shifts are shown in Fig.\ref{fig22}. Here we see for the individual orbital momenta quite pronounced oscillations, but the subtracted phase shift is much smoother.

\begin{figure}[h]
\begin{center}
\includegraphics{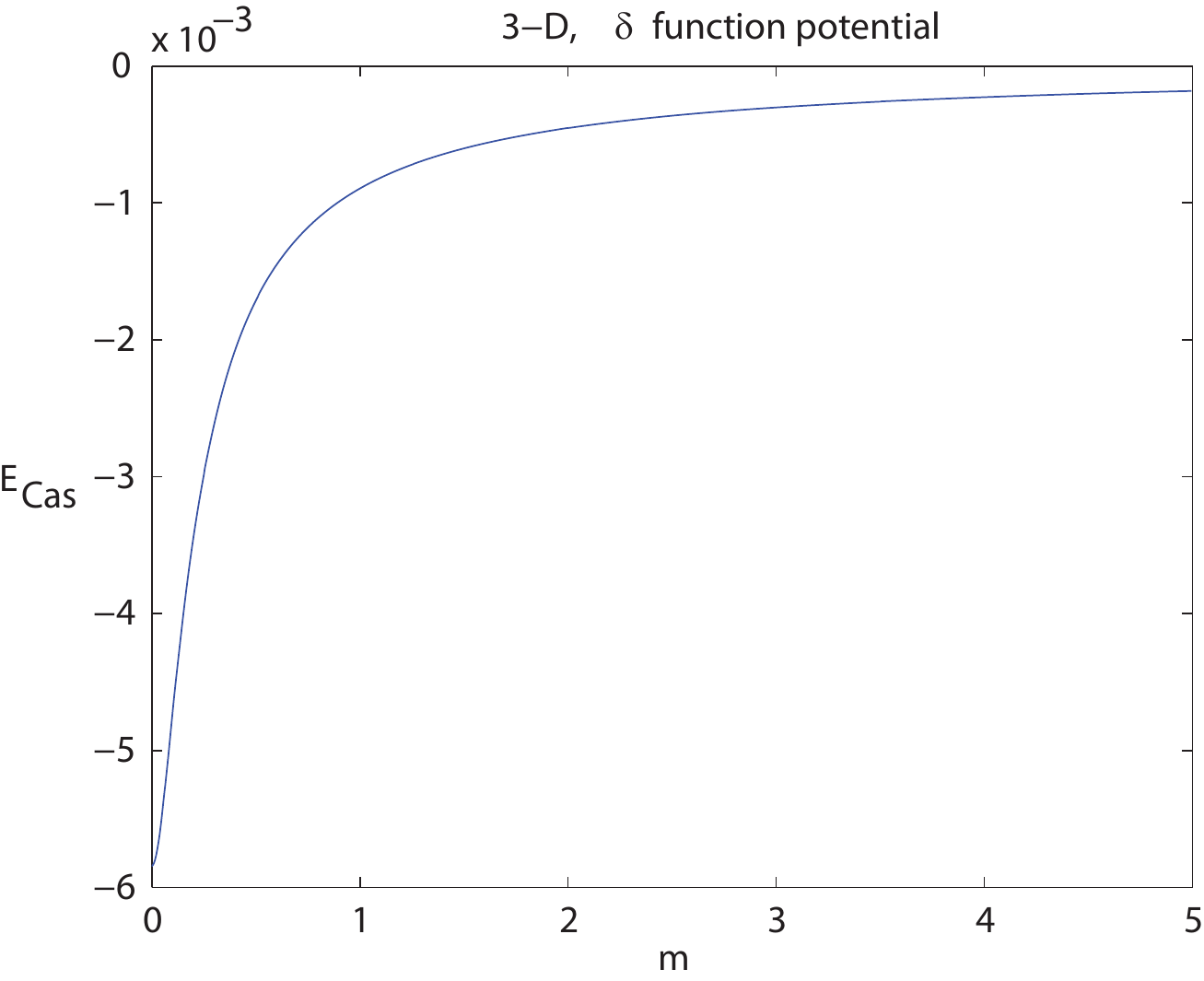}
\caption{The Casimir energy for the 3-dimensional case for delta function potentials for $R=1$ and $\alpha =1$.}
\label{fig2}
\end{center}
\end{figure}

Use of the eqs. (\ref{37}) and (\ref{38}) in (\ref{23}) gives the results for the Casimir energy shown in Fig. \ref{fig2} for $R=1$ and $\alpha =1$.

\section{Casimir energy for several examples in two dimensions}
\subsection{Step potential}
\begin{figure}[h]
\begin{center}
\includegraphics{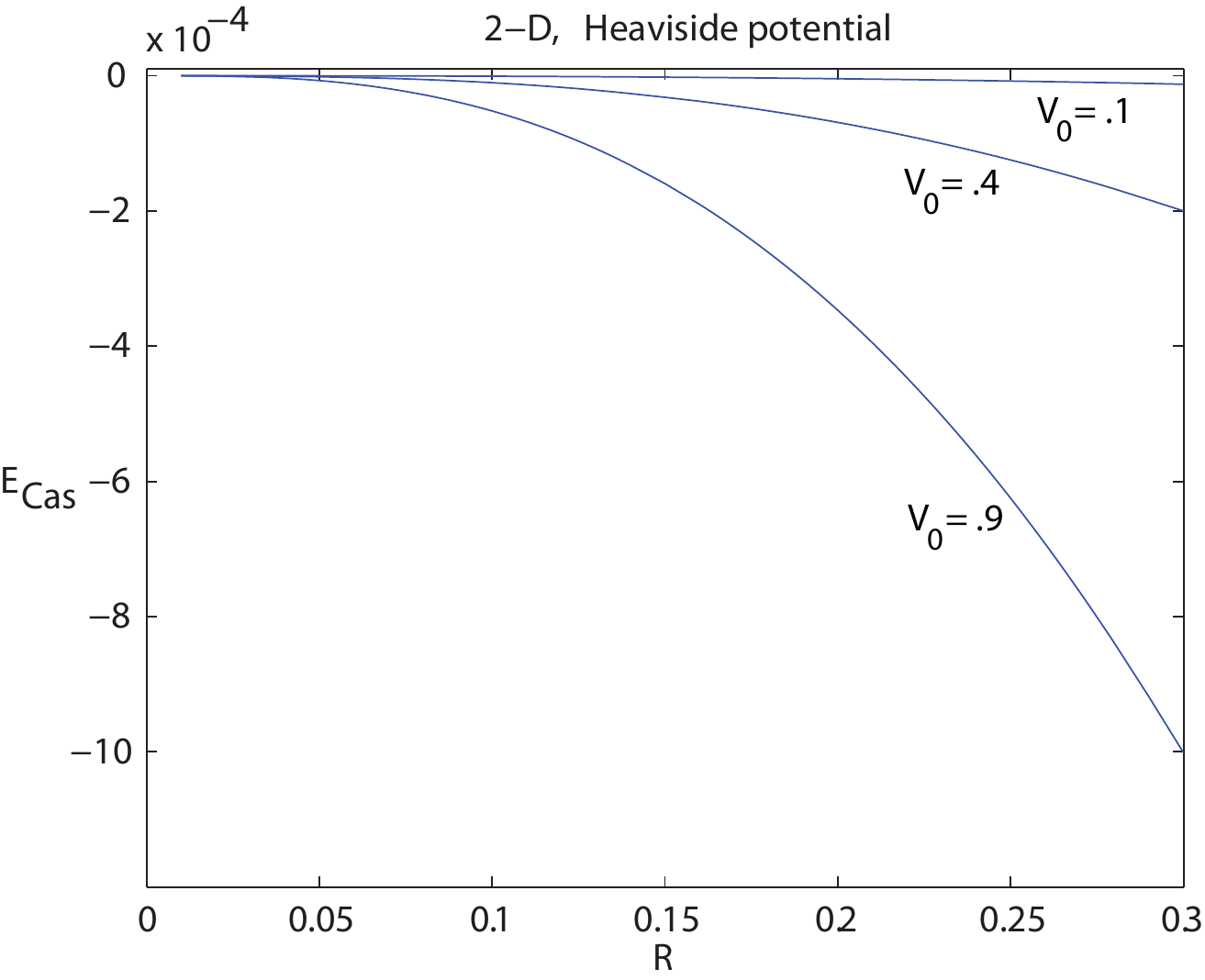}
\caption{The Casimir energy for the 2-dimensional case for heaviside potentials for $m=1$.}
\label{fig3}
\end{center}
\end{figure}

The computations in two dimensions mimic very much the just presented analysis. As is clear from the eigenvalue eqs. (\ref{8})
and (\ref{19}), many results in two dimensions follow from the corresponding three dimensional ones by replacing $\ell$ with $m-1/2$.
From this analogy, the phase shift for a two-dimensional step potential reads
\beq
\delta_m (p) = - \arctan \frac{ q J_m (pR) J_m ' (qR) - p J_m (qR) J_m ' (pR)}{p J_m (qR) N_m ' (pR) - q N_m (pR) J_m ' (qR) } . \label{39}
\eeq
The needed heat kernel coefficients are
\beq
a_0  = a_{1/2} = a_{3/2} =0, \quad a_1 = - \frac 1 4 V_0 R^2, \label{40}
\eeq
and from eq. (\ref{23}) the Casimir energy as given in Fig. \ref{fig3} for $m=1$ is found.

\subsection{Delta function potential}
\begin{figure}[h]
\begin{center}
\includegraphics{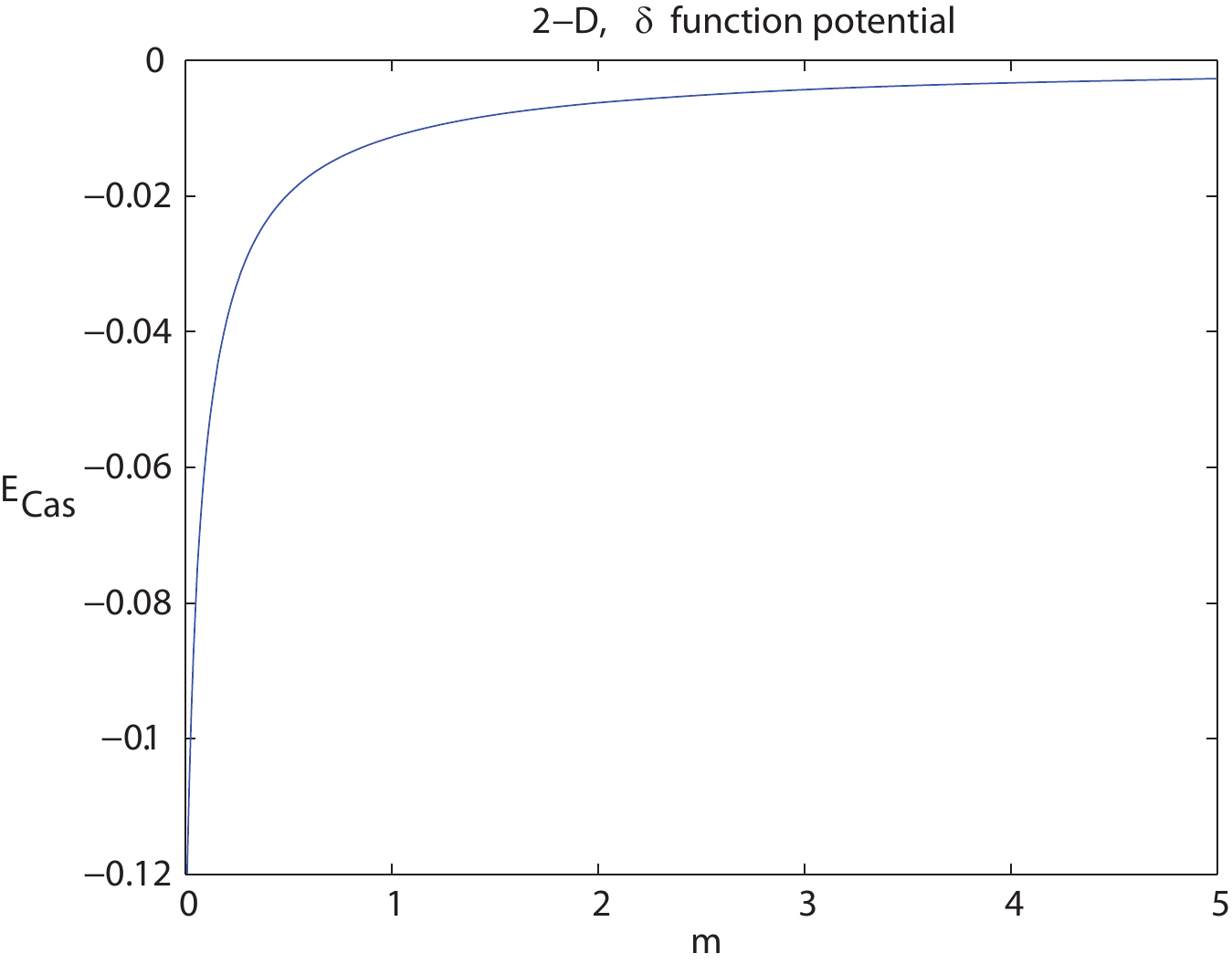}
\caption{The Casimir energy for the 2-dimensional case for delta function potentials for $R=1$ and $\alpha 0.88$.}
\label{fig4}
\end{center}
\end{figure}

With the same analogy as before, $\ell \to m-1/2$, the phase shift is from (\ref{37})
\beq
\delta_m (p) = - \arctan \frac{ \frac{\pi \alpha} 2 J_m^2 (pR)} {1-\frac {\pi \alpha} 2 J_m (pR) N_m (kR) } . \label{41}
\eeq
The heat kernel coefficients are
\beq
a_0 = a_{1/2} =0, \quad a_1 = - \frac \alpha 2 , \quad a_{3/2} = \frac{ \sqrt \pi} 8 \frac{\alpha^2} R, \label{42}
\eeq
and again from (\ref{23}) the Casimir energy follows; see Fig. \ref{fig4} for $\alpha = 0.88$.

\section{Conclusions}
In this article we have provided a much improved formalism for the computation of vacuum energies in the presence of spherically symmetric potentials.
The central result is in eqs. \Ref{18_0} and \Ref{18}  which express the renormalized Casimir energy in terms of the subtracted phase shift \Ref{18}  appearing after summation over the orbital momenta and subsequent subtraction of contributions given in terms of the heat kernel coefficients. These are known for a wide range of potentials and they are easily computed \cite{gilk95b}.
The remaining task thus is to compute $\delta (q)$ and some integral numerically.

For examples where $\delta_\ell (q)$ is known analytically the numerical computation can be done based on that knowledge. This is what we have done here for
step and delta potentials in two and three dimensions.

Typically, $\delta _\ell (q)$ will not be known analytically, in which case the probably most suitable way for its evaluation is the variable phase approach \cite{tayl72b}.
We envisage to treat more complicated examples along these lines in a separate publication.

Furthermore, clearly our approach can be used for spinors or other fields. In this case the heat kernel coefficients and $\delta (q)$ will contain traces over internal degrees,
but that represents a very minor complication. The procedure again should allow for a considerably simplified analysis of for example flux tubes \cite{bord99-60-105019,grah05-707-233}
and other configurations.



\end{document}